\documentclass[lettersize,journal]{IEEEtran}
\usepackage{amsmath,amsfonts}
\usepackage{algorithmic}
\usepackage[ruled,noline]{algorithm2e}
\usepackage{array}
\usepackage{tabularx}
\usepackage{makecell}  
\usepackage{textcomp}
\usepackage[caption=true,font=footnotesize,textfont=rm]{subfig}
\usepackage{subfloat}
\usepackage{textcomp}
\usepackage{tikz}

\usepackage{amssymb}
\usepackage{bbding}
\usepackage{pifont}
\usepackage{url}
\usepackage{fancyhdr}
\usepackage{hyperref}
\usepackage{verbatim}
\usepackage{graphicx}
\usepackage{multirow}
\usepackage{cite}
\usepackage{booktabs}
\usepackage{ulem}

\usepackage{float}
\usepackage{makecell}
\usepackage{orcidlink}
\usepackage{tikz}
\usepackage[misc]{ifsym}
\usepackage[switch]{lineno}
\begin{document}
\title{DiM-Gesture: Co-Speech Gesture Generation with Adaptive Layer Normalization Mamba-2 framework}

\author{Fan Zhang\hspace{-1.5mm}$^{~\orcidlink{0000-0002-9534-1777}}$,
Naye Ji $^{(\textrm{\Letter})}$ \hspace{-1.5mm}$^{~\orcidlink{0000-0002-6986-3766}}$,
Fuxing Gao\hspace{-1.5mm}$^{~\orcidlink{0009-0008-5586-4734}}$,
Bozuo Zhao \hspace{-1.5mm}$^{~\orcidlink{0009-0008-2152-0087}}$,
Jingmei Wu \hspace{-1.5mm}$^{~\orcidlink{0009-0000-9221-8784}}$,
Yanbing Jiang \hspace{-1.5mm}$^{~\orcidlink{0009-0003-3020-598X}}$,
Hui Du\hspace{-1.5mm}$^{~\orcidlink{0000-0001-6737-9420}}$, 
Zhenqing Ye\hspace{-1.5mm}$^{~\orcidlink{0009-0003-4341-2734}}$, 
Leyao Yan \hspace{-1.5mm}$^{~\orcidlink{0009-0008-8797-176}}$,
Jiayang Zhu \hspace{-1.5mm}$^{~\orcidlink{0009-0005-3072-651X}}$,
WeiFan Zhong\hspace{-1.5mm}$^{~\orcidlink{0009-0006-3251-1392}}$,
Xiaomeng Ma \hspace{-1.5mm}$^{~\orcidlink{0009-0008-0112-5354}}$ %

\thanks{Fan Zhang, Naye Ji, Zhenqing Ye, Jiayang Zhu, WeiFan Zhong, Leyao Yan, Yanbing Jiang, Hui Du, Fuxing Gao, are with the College of Media Engineering, Communication University of Zhejiang, China; (e-mail: fanzhang@cuz.edu.cn;jinaye@cuz.edu.cn; 1767641184@qq.com; zjy190127 26077@foxmail.com; 2633252395@qq.com; yanleyao8@gmail.com; duhui@ cuz.edu.cn; fuxing@cuz.edu.cn;)}

\thanks{Jingmei Wu, Xiaomeng Ma, are with the School of Broadcast Announcing Arts, Communication University of Zhejiang; (e-mail: 20190095@cuz.edu.cn; 1256914539@qq.com)}
\thanks{Zhao Bozuo is with Changjiang Academy of Art and Design, Shantou University, China (e-mail: bzzhao@stu.edu.cn)}
}




\maketitle

\begin{abstract}
Speech-driven gesture generation is an emerging domain within virtual human creation, where current methods predominantly utilize Transformer-based architectures that necessitate extensive memory and are characterized by slow inference speeds. In response to these limitations, we propose \textit{DiM-Gestures}, a novel end-to-end generative model crafted to create highly personalized 3D full-body gestures solely from raw speech audio, employing Mamba-based architectures. This model integrates a Mamba-based fuzzy feature extractor with a non-autoregressive Adaptive Layer Normalization (AdaLN) Mamba-2 diffusion architecture. The extractor, leveraging a Mamba framework and a WavLM pre-trained model, autonomously derives implicit, continuous fuzzy features, which are then unified into a singular latent feature. This feature is processed by the AdaLN Mamba-2, which implements a uniform conditional mechanism across all tokens to robustly model the interplay between the fuzzy features and the resultant gesture sequence. This innovative approach guarantees high fidelity in gesture-speech synchronization while maintaining the naturalness of the gestures. Employing a diffusion model for training and inference, our framework has undergone extensive subjective and objective evaluations on the ZEGGS and BEAT datasets. These assessments substantiate our model's enhanced performance relative to contemporary state-of-the-art methods, demonstrating competitive outcomes with the DiTs architecture (Persona-Gestors) while optimizing memory usage and accelerating inference speed. Code can be accessed at \href{https://github.com/zf223669/DiMGestures} {https://github.com/zf223669/DiMGestures.}

\end{abstract}


\section{Introduction}
\IEEEPARstart{R}{ecent} advancements have significantly broadened the scope of 3D virtual human technology, with its applications permeating diverse fields such as animation, gaming, human-computer interaction, and digital reception. Central to this research is the development of credible, personalized co-speech gestures. Speech-driven gesture generation, facilitated by deep learning, offers an efficient alternative to traditional motion capture systems, which typically require extensive manual input.

However, a fundamental challenge within this domain is the identification and integration of numerous input conditions essential for effective gesture synthesis. This complexity stems from a variety of factors, such as acoustic nuances, semantic content, emotional expressions, personality traits, and demographic elements like gender and age. These components collectively influence the dynamics of co-speech gestures, necessitating sophisticated models that can interpret and synthesize naturalistic human movements based on verbal communication.

Previous approaches\cite{bhattacharya2021speech2affectivegestures,yang_diffusestylegesture_2023,yang2023DiffuseStyleGestureaplus,alexanderson2023listen, yang_unifiedgesture_2023, li2023Audio2Gesturesa,ghorbani2023ZeroEGGS} have employed manual labels and diverse feature inputs to facilitate the synthesis of personalized gestures. However, these methods rely heavily on varied unstructured feature inputs and necessitate complex multimodal processing, posing significant barriers to the practical application and broader adoption of virtual human technologies.

The fuzzy inference strategy, derived from the concept of fuzzy logic\cite{zadeh1965fuzzy}, has proven particularly useful in fields requiring the management of uncertain or imprecise information. This approach is noted for its effectiveness in applications such as speech-emotion recognition\cite{vashishtha2020unsupervised} and audio classification\cite{patil2019content}. Unlike methods that rely on explicit classification outputs, fuzzy inference provides a spectrum of feature information, transitioning from a limited explicit discrete space to an expansive implicit continuous fuzzy space. This fuzzy space aligns more closely with real-world scenarios. Psychological research underscores the importance of various factors in speech\cite{calvo2015oxford,goudbeek2010beyond,hirschberg2015advances,campbell2011intersecting}, which, when considered as fuzzy features, are intricately linked to co-speech gestures. Zhang et al.\cite{zhang_speech-driven_2024} pioneered the successful integration of fuzzy inference strategies for generating personalized co-speech gestures.

Achieving high gesture-speech synchronization while maintaining naturalness poses a significant challenge in the field of speech-driven gesture generation. Recent advancements have pivoted towards employing Transformer and Diffusion-based models, enhancing the efficiency and flexibility of gesture-generation technologies. Notable innovations in this domain include Diffuse Style Gesture\cite{yang_diffusestylegesture_2023}, Diffuse Style Gesture+\cite{yang2023DiffuseStyleGestureaplus}, GestureDiffuClip\cite{ao2023GestureDiffuCLIP}, and LDA\cite{alexanderson2023listen}. However, these methods sometimes struggle with maintaining an optimal correlation between gesture and speech, which can detract from the naturalness of the gestures produced.

The emergence of Diffusion Transformers (DiTs) in fields such as text-to-image generation \cite{peebles2022Scalable} and text-to-video tasks like Sora\footnote{https://openai.com/sora}, which utilize Adaptive Layer Normalization (AdaLN), represents a substantial leap forward. These models introduce a uniform conditional mechanism across all tokens, significantly improving the representation of conditional and output features. This enhancement is poised to effectively model the complex relationship between speech and gestures. Despite the Persona-Gestor \cite{zhang_speech-driven_2024} architecture's ability to generate high-quality co-speech gestures using DiT, it faces challenges such as high memory demands and slower inference speeds due to its convolutional and transformer-based structure. 

The Mamba-bases frameworks \cite{gu_mamba_2023,dao_transformers_2024}, a relatively new entrant, has garnered attention for its simplicity, efficiency, and flexibility. These findings also underscore the Mamba architecture as a formidable competitor to the Transformer architecture, offering compelling alternatives in handling complex tasks efficiently. 

In this research, we introduce \textit{DiM-Gesture}, a groundbreaking model designed to synthesize personalized gestures exclusively from raw speech audio. This approach employs a fuzzy feature inference strategy within its condition extractor, utilizing the Mamba architecture, and integrates an AdaLN Mamba-2 in a diffusion-based module. \textit{DiM-Gesture} transitions from relying on explicit conditions to a refined, continuous representation of fuzzy features, effectively capturing a wide range of stylistic nuances and specific audio details. These features are amalgamated into a unified latent representation, facilitating the synthesis of intricate 3D full-body gestures. The implementation of AdaLN significantly bolsters the model's ability to articulate the complex relationship between speech and gestures. Leveraging a diffusion process, the model is capable of generating diverse, high-fidelity gesture outputs, showcasing its effectiveness in gesture synthesis.

For clarity, our contributions are summarized as follows:
\begin{itemize}
\item{\textbf{Introduction of a pioneering infer Mamba-based fuzzy features strategy:}} This strategy allows the synthesis of a broader range of personalized gestures solely from speech audio, eliminating the need for style labels or additional inputs. The Mamba-based fuzzy feature extractor significantly enhances the system's usability capabilities.

\item{\textbf{Integration of the AdaLN Mamba-2 architecture within the diffusion model:}} This architectural choice improves the modeling of the intricate interplay between speech and gestures. Our findings validate that the Mamba architecture is a formidable competitor to traditional Transformer architectures, demonstrating an optimal balance of naturalness and synchronization in gesture generation.

\item{\textbf{Extensive subjective and objective evaluations:}} These evaluations confirm that our model surpasses current state-of-the-art approaches, highlighting the exceptional capability of our method to generate credible, speech-appropriate, and personalized gestures.
\end{itemize}

\section{RELATED WORK}\label{sec:RELATED WORK}
The discussion presented provides a concise overview of Transformer-based and diffusion-based generative models within the realm of speech-driven gesture generation.

DiffMotion \cite{zhang2023diffmotion} marks an innovative use of diffusion models in gesture synthesis, integrating an LSTM to generate diverse gestures.  Alexanderson et al. \cite{alexanderson2023listen} have refined DiffWave by replacing dilated convolutions to optimize the potential of transformer architectures. Conformers \cite{gulati2020Conformer} employ classifier-free guidance to enhance style expression. GestureDiffuCLIP \cite{ao2023GestureDiffuCLIP} utilizes transformers and AdaIN layers to integrate style guidance within the diffusion process. DiffuseStyleGesture (DSG) \cite{yang_diffusestylegesture_2023} and its extension DSG+ \cite{yang2023DiffuseStyleGestureaplus} incorporate cross-local attention and layer normalization within their transformer models. However, these methodologies often face challenges in striking an optimal balance between gesture and speech synchronization, potentially leading to gestures that seem either underrepresented or excessively matched to the speech. Persona-Gestor \cite{zhang_speech-driven_2024} introduces a fuzzy feature extractor that leverages a 1D convolution to process raw speech audio for feature extraction, combined with an Adaptive Layer Normalization (AdaLN) transformer \cite{peebles2022Scalable} to model the intricate correlation between these features and the resulting gesture sequence. While this method achieves superior motion quality, the significant memory requirements and slower inference speeds of convolutional and transformer architectures remain a challenge.

In this study, we implement a fuzzy feature inference strategy to implicitly capture nuanced features within speech audio. We synthesize natural, personalized co-speech gestures exclusively based on raw speech audio, leveraging the Mamba architecture. Furthermore, we employ the AdaLN Mamba-2 architecture in place of traditional AdaLN transformers. This adaptation preserves the model’s ability to capture the complex interplay between speech and gestures while ensuring the generation of high-quality actions. Notably, this approach significantly reduces memory requirements and enhances inference speed, offering a more efficient alternative for gesture synthesis in virtual human interactions. 
\begin{figure*}[!ht]
\centering
\subfloat[Overall schematic]{\includegraphics[width=0.34\textwidth]{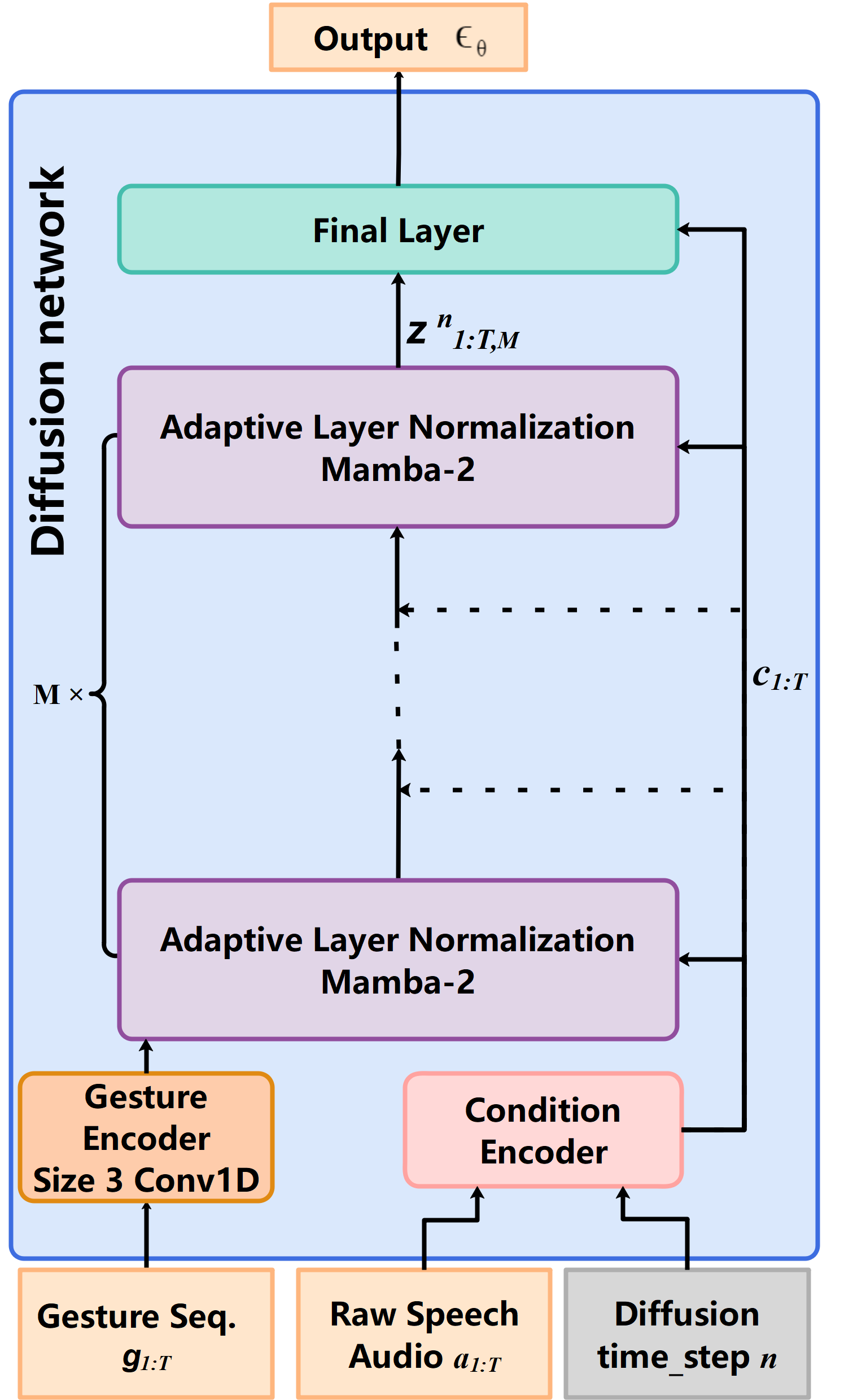}%
\label{fig:architecture1}}
\hfil
\subfloat[Fuzzy Feature Extractor (below), Final Layer in AdaLN Transformer (above)]{\includegraphics[width=0.33\textwidth]{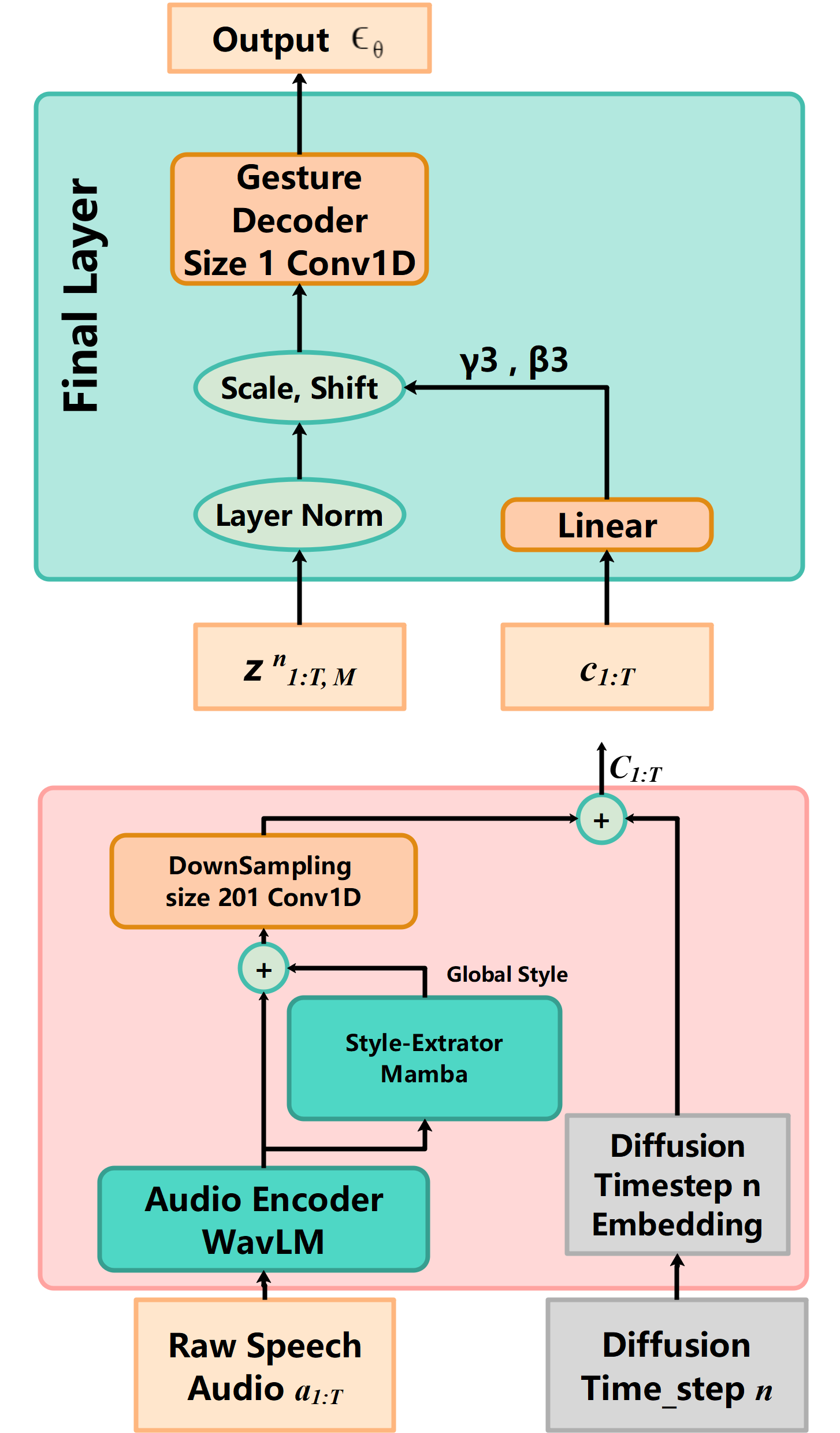}%
\label{fig:architecture2}}
\hfil
\subfloat[AdaLN Transformer Block]{\includegraphics[width=0.311\textwidth]{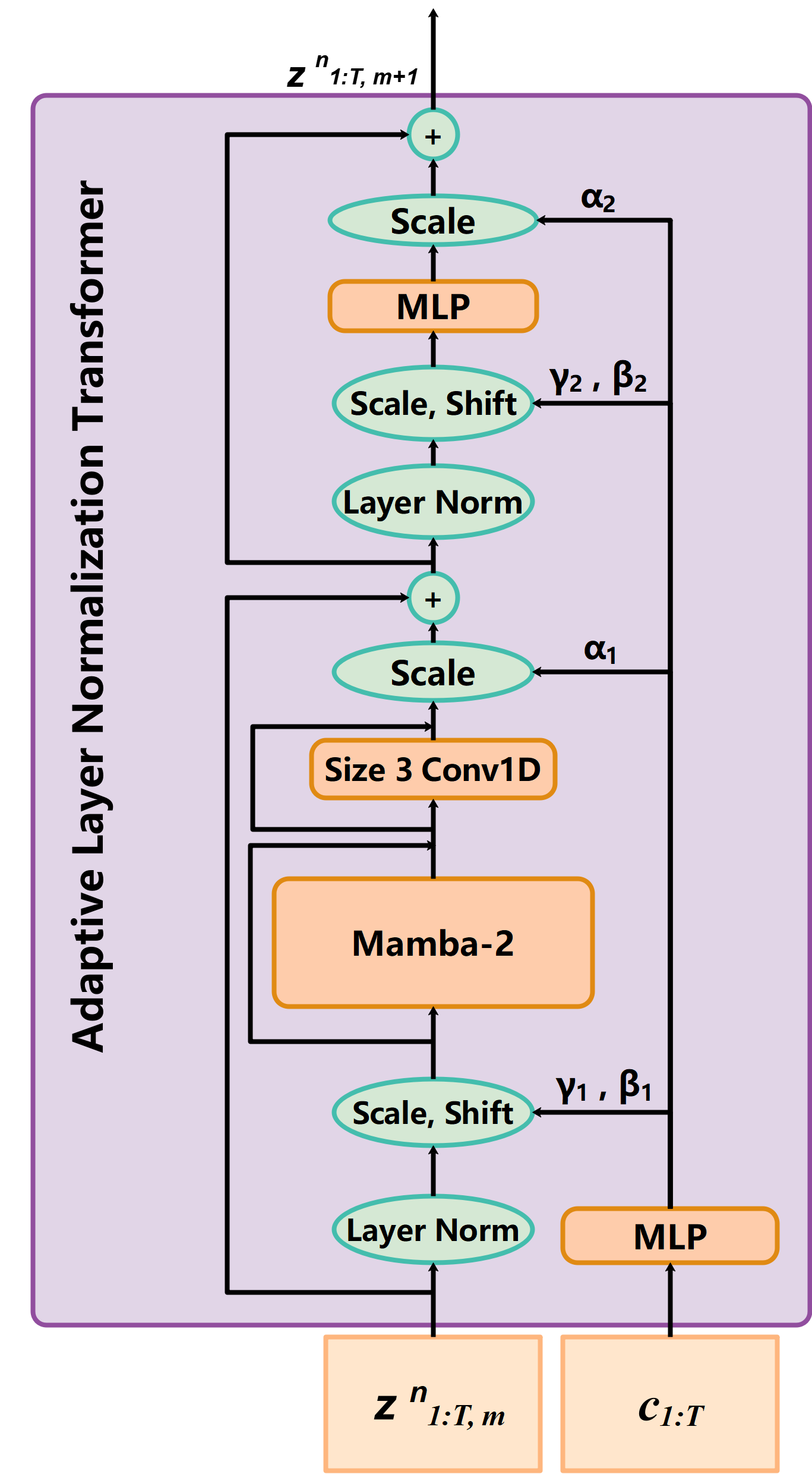}%
\label{fig:architecture3}}
\caption{The architecture of DiM-Gesture strategically incorporates a Mamba-based fuzzy feature extractor and an Adaptive Layer Normalization (AdaLN) Mamba-2 diffusion architecture. At its core, the fuzzy feature extractor features a dual-component system engineered to capture both the nuanced style and detailed audio features. These unified latent features are subsequently channeled into the AdaLN Mamba-2. This module is pivotal in modeling the intricate relationship between the incoming audio features and the corresponding gestures. It facilitates the precise estimation of diffusion noise within the diffusion model, ensuring the generation of diverse gestures. The overall schematic includes three main components: (a) Overall Schematic, (b) Mamba-based Style Fuzzy Feature Extractor, and (c) AdaLN Mamba-2 Block.}
\label{fig:architecture}
\end{figure*}

\section{System Overview}\label{sec:PROPOSED APPROACH}
DiM-Gesture, an end-to-end Mamba and diffusion-based architecture, processes raw speech audio as its sole input to synthesize personalized gestures. This model adeptly balances naturalness with synchronized alignment to the corresponding speech, ensuring the gestures it generates are both realistic and timely, enhancing the interaction experience in virtual environments. 

\subsection{Problem Formulation}
In this study, we frame the problem of co-speech gesture generation as a sequence-to-sequence translation challenge, where the objective is to map a sequence of speech audio features, denoted as \(a = a_{1:T} = [a_1, \ldots, a_t, \ldots, a_T] \in \mathbb{R}^{T}\), to a corresponding sequence of full-body gesture features, represented as \(g^0 = g^0_{1:T} = [g^0_1, \ldots, g^0_t, \ldots, g^0_T] \in \mathbb{R}^{T \times (D+6)}\). Here, each \(g^0_t \in \mathbb{R}^{(D+6)}\) comprises 3D joint angles alongside root positional and rotational velocities at frame \(t\), with \(D\) indicating the number of joint channels.

We define the probability density function (PDF), \(p_\theta(\cdot)\), to approximate the actual gesture data distribution \(p(\cdot)\), facilitating efficient sampling. Our aim is to generate a non-autoregressive whole pose sequence (\(g^0\)) from its conditional probability distribution given the audio sequence (\(a\)) as a covariate:

\[
g^0 \sim p_\theta(g^0 \mid a) \approx p(\cdot) := p(g^0 \mid a)
\]

This formulation underscores the use of a denoising diffusion model trained to approximate the true conditional distribution of gestures given speech, illustrating the direct relationship between the audio inputs and gesture outputs, thus setting the stage for our model to learn this complex mapping effectively.

\subsection{Model Architecture}
The architecture of DiM-Gesture, illustrated in Figure \ref{fig:architecture}, is composed of four main components designed to streamline the synthesis of personalized gestures from speech audio. These components include:

1) Mamba-based Style Fuzzy Feature Extractor, 2) AdaLN Mamba-2 Architecture, 3) Gesture Encoder and Decoder, and 4) Diffusion Network.

Together, these components form a cohesive system that not only captures the complexity of human gestures relative to speech but also enhances the quality and personalization of the generated gestures.

\subsubsection{Mamba-based Fuzzy Feature Extractor}

\begin{figure}[htbp]
\includegraphics[width=0.45\textwidth]{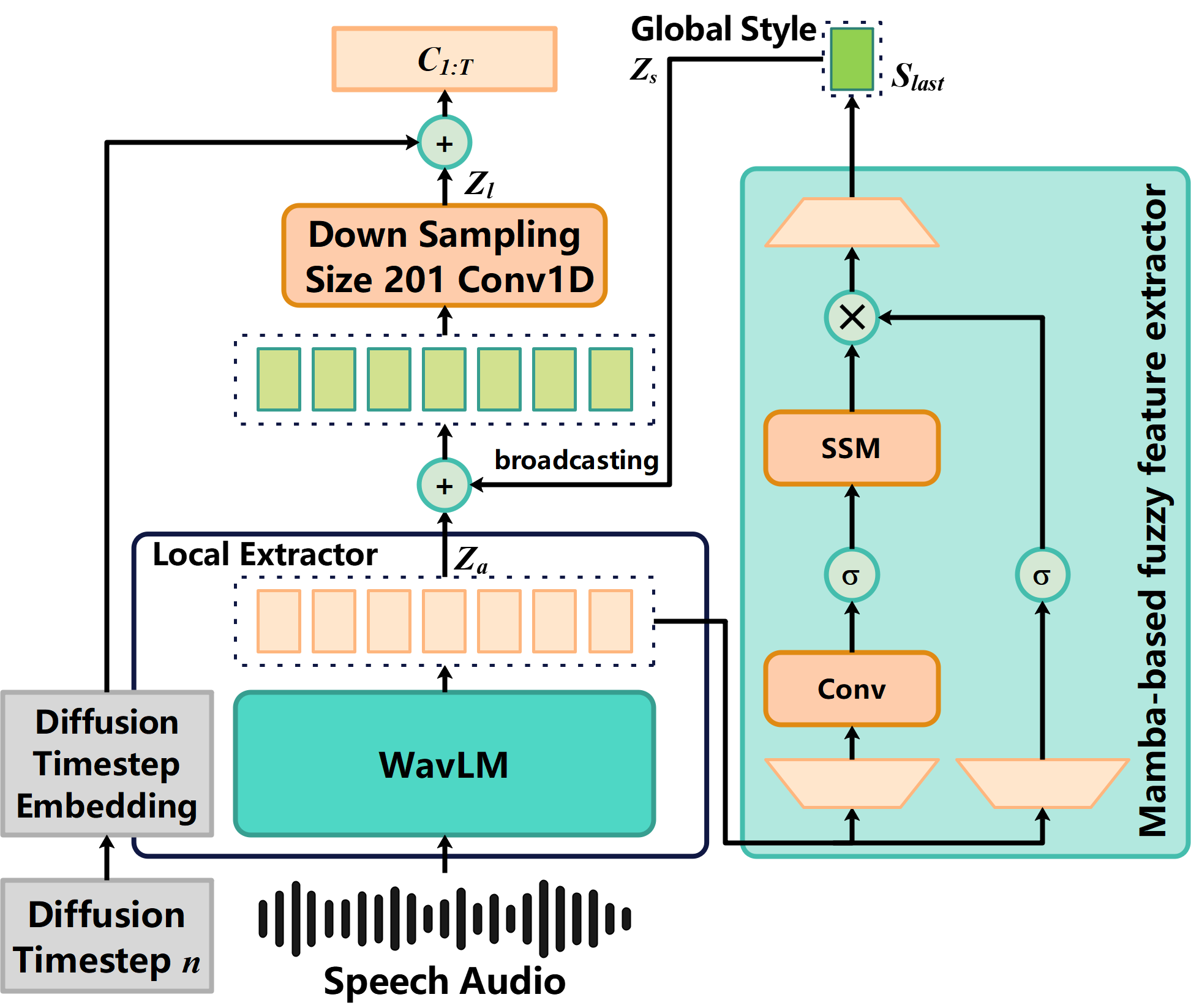}
\begin{center}
\caption{An overview of the Mamba-based style fuzzy inference condition extractor. }
\label{fig:ConditionEncoder} 
\end{center}
\end{figure}

This module employs a fuzzy inference strategy, which means it does not rely on explicit classification outputs. Instead, it provides implicit, continuous, and fuzzy feature information, automatically learning and inferring the global style and specific details directly from raw speech audio. Illustrated in Figure \ref{fig:architecture2} and Figure \ref{fig:ConditionEncoder}, this module is a dual-component extractor comprising both global and local extractors. The local extractor utilizes the WavLM large-scale pre-trained model\cite{chen2022Wavlm} to process the audio sequence into discrete tokens. We selected WavLM for its proficiency in capturing the complex attributes of speech audio, which allows it to effectively represent universal audio latent features, denoted as $z_a$. This sophisticated approach ensures a nuanced translation of audio features into gesture-relevant data, enhancing the fidelity and personalization of the generated gestures.

We implement a global style extractor within our system, utilizing the Mamba module \cite{gu_mamba_2023} to process $z_a$, the universal audio latent representations. This global extractor is adept at automatically capturing and embedding global fuzzy style information from $z_a$, producing an output token $z_{last} \in \mathbb{R}^{1\times D'}$, which is the last output of the Selective State Space Model(SSM). This token is subsequently broadcasted and amalgamated with $z_a \in \mathbb{R}^{T'\times D'}$ to forge a unified latent representation $z_l \in \mathbb{R}^{T\times D''}$. By merging both local and global insights, our architecture enhances the representational fidelity of the entire sequence, tailored for co-speech gesture generation. This unified latent representation is then channeled to the downsampling module for further refinement. 

The downsampling module, crucial for aligning each latent representation with its corresponding sequence of encoded gestures, is seamlessly integrated into the condition extractor. Initially, we explored linear alignment strategies, similar to those employed in DSG\cite{yang_diffusestylegesture_2023} and DSG+\cite{yang2023DiffuseStyleGestureaplus}. However, these methods often led to issues such as foot-skating. To address this, we have implemented a Conv1D layer with a kernel size of 201 within our architecture. This configuration enables the mapping of every 201-length target token output from the WavLM to a single gesture frame, thereby enhancing the precision of gesture synchronization. The output of this module, $c_{1}$, serves as a unified latent representation that encapsulates both encoded audio features and the diffusion time step $n$, ensuring a coherent and accurate gesture generation process. 

\subsubsection{AdaLN Mamba-2}
The AdaLN's fundamental purpose is to incorporate a conditional mechanism that uniformly applies a specific function across all tokens, thereby significantly improving the model's capacity for representing both conditional and output features with enhanced efficiency. It offers a more sophisticated and nuanced approach to modeling, enabling the system to capture and articulate the complex dynamics between various input conditions and their corresponding outputs. Consequently, this leads to an improvement in the model's predictive accuracy and ability to generate outputs more aligned with the given conditions.  

Diffusion Transformers (DiTs) exemplify a sophisticated advancement in diffusion model architectures, incorporating an AdaLN-infused transformer framework primarily for text-to-image synthesis. This methodological enhancement has substantially lowered Fréchet Inception Distance (FID) scores, demonstrating improved image quality. The utility of DiTs has recently expanded to include text-conditional video generation, illustrating their versatility. Furthermore, DiTs have shown potential in co-speech gesture generation \cite{zhang_speech-driven_2024}, marking a significant step in applying these models to sequence-based tasks. However, the inherent quadratic space complexity associated with Transformers results in substantial memory consumption and slower inference speeds, presenting practical challenges in real-time applications. 

In contrast to traditional Diffusion Transformers, our approach integrates the Mamba-2 architecture \cite{dao_transformers_2024} as a replacement for the conventional transformer module. This strategic adaptation leverages the minimal memory footprint and enhanced processing efficiency of Mamba-2, significantly accelerating inference speeds without sacrificing output quality. This novel substitution is pivotal in addressing the challenges associated with real-time, speech-driven gesture synthesis, ensuring efficient and high-quality performance.

The module involves regressing the dimension-wise scale and shift parameters (\(\gamma\) and \(\beta\)), derived from the fuzzy feature extractor output \(c_{1:T}\), rather than directly learning these parameters, as depicted in Figure \ref{fig:architecture3}. In each AdaLN Mamba-2 stack, a latent feature \(z^n_{1:T,m}\) is generated, combining condition information and gesture using AdaLN and the Mamba-2 architecture. The index \(m\) ranges from 1 to \(M\), where \(M\) represents the total number of AdaLN Mamba-2 stacks. Furthermore, as illustrated in Figure \ref{fig:architecture2}, the final layer utilizes the same fuzzy features, supplemented by a scale and shift operation to fine-tune the gesture synthesis.

This method enables the creation of detailed gesture sequences directly from speech audio, obviating the need for discrete style labels or additional inputs. It significantly enhances the model's ability to generate personalized and contextually aligned gestures, providing a refined and context-sensitive gesture synthesis capability. Moreover, compared with the AdaLN transformer, the Mamba-2 architecture reduces the memory footprint and improves inference speed, while still ensuring high generation quality. 

\subsubsection{Gesture Encoder and Decoder}
The architecture of the gesture encoder and decoder is designed to process the gesture sequence, as illustrated in Fig. \ref{fig:architecture1} and Fig. \ref{fig:architecture2}. The gesture encoder employs a Convolution1D layer with a kernel size of 3 to encode the initial sequence of gestures \(g\) into a hidden state \(h \in \mathbb{R}^{T \times D''}\). Our experimental results indicate that a kernel size of 1 tends to produce animation jitter. In contrast, a kernel size of 3 effectively mitigates this issue by capturing the spatial-temporal relationships inherent in gesture sequences.

The gesture decoder then reduces the feature dimension of the transformer's output \(D''\) back to the original dimension \(D\), which corresponds to the number of channels representing skeleton joints, resulting in the output of the predicted noise (\(\epsilon_\theta\)). Employing a 1D kernel of size 1 in the input sequence allows the model to effectively extract meaningful features and relationships between adjacent joint channels, enhancing the quality and coherence of the generated gestures.


\subsection{Training and Inferencing with DDPM} \label{sec:DDPM}
The diffusion process in our architecture is crucial for reconstructing the conditional probability distribution between gestures and fuzzy features. This involves employing a systematic sampling approach from this reconstructed distribution, facilitating the generation of diverse gestures.

Following our previous work, Persona-Gestor \cite{zhang_speech-driven_2024}, incorporating the Denoising Diffusion Probabilistic Model (DDPM) into our approach. 

The model operates through two principal processes: the diffusion process and the generation process. In the diffusion process during training, the model gradually transforms the original gesture data (\(g^0\)) into white noise (\(g^N\)) by optimizing a variational bound on the data likelihood. This transformation is characterized by progressively adding noise to the data in a controlled manner.

During inference, the generation process endeavors to reverse this transformation. It recovers the original data from the noise by reversing the noising process through a Markov chain, employing Langevin sampling \cite{paul_sur_1908}. This technique ensures the effective and accurate reconstruction of the gesture data from its noised state. The Markov chains utilized in both the diffusion and generation processes ensure a coherent and systematic transition between stages, thereby maintaining the integrity and quality of the generated gestures. This dual-process framework allows the model to efficiently handle and synthesize complex gesture data, reflecting the dynamic nature of human movement. The Markov chains in the diffusion process and the generation process are:
\begin{equation}
\begin{aligned}
&p\left(g^n|g^0\right) = \mathcal{N}\left(g^n; \sqrt{\overline{\alpha}^n} g^0, \left(1-\overline{\alpha}^n\right)I\right)   \quad and\\ 
&p_\theta\left(g^{n-1}|g^n, g^0\right) = \mathcal{N}\left(g^{n-1}; \tilde{\mu}^n\left(g^n, g^0\right), \tilde{\beta}^n I\right),
\end{aligned}
\label{eq:cumulativeProduct}
\end{equation}
where \(\alpha^n := 1 - \beta^n\) and \(\overline{\alpha}^n := \prod_{i=1}^n \alpha^i\). As shown by\cite{ho_denoising_2020}, \(\beta^n\) is a increasing variance schedule \(\beta^1,...,\beta^N\) with \(\beta^n \in (0,1)\), and \(\tilde{\beta}^n := \frac{1-\overline{\alpha}^{n-1}}{1-\overline{\alpha}^n}\beta^n\).

The training objective centers on optimizing the parameters \(\theta\) by minimizing the Negative Log-Likelihood (NLL). This is achieved via the Mean Squared Error (MSE) loss, which quantifies the discrepancy between the true noise, represented by \(\epsilon \sim \mathcal{N}(0, I)\), and the predicted noise \(\epsilon_\theta\):

\begin{equation}
\label{eq:objective2}
\mathbb{E}_{g^0_{1:T}, \epsilon, n}[||\epsilon - \epsilon_\theta\left(\sqrt{\overline{\alpha}^n g^0}+\sqrt{1-\overline{\alpha}^n}\epsilon , a_{1:T},n\right)||^2],
\end{equation}  Here \(\epsilon_\theta\) is a neural network (see figure \ref{fig:architecture1}), which uses input \(g_t^0\) , \(a_{t-1}\) and \(n\) that to predict the \(\epsilon\), and contains the similar architecture employed in \cite{rasul_autoregressive_2021}. The complete training procedure is outlined in Algorithm \ref{alg:Training}.

\begin{algorithm}[htbp]
    \caption{Training for the whole sequence gesture}
    \KwIn{data $g^0_{1:T} \sim p\left(g^0|a_{1:T}\right)$ and \(a_{1:T}\)}
    \Repeat{converged}{Initialize $n \sim$ Uniform$(1,...,N)$ and $\epsilon \sim \mathcal{N}(0,I)$
    \\Take the gradient step on
    $$\nabla_\theta||\epsilon-\epsilon_\theta\left(\sqrt{\overline{\alpha}_n}g^0_{1:T}+\sqrt{1-\overline{\alpha}_n}\epsilon, a_{1:T},n\right)||^2$$}    \label{alg:Training}
\end{algorithm}

After the training phase, we employ variational inference to generate a complete sequence of new gestures that match the original data distribution (\(g^0 \sim p_\theta(g^0 \mid a)\)). This is facilitated by the sampling procedure detailed in Algorithm \ref{alg:Inference}. During this process, we sample the entire sequence \(g^0\) from the learned distribution.

The term \(\sigma_\theta\) represents the standard deviation of the conditional distribution \(p_\theta(g^{n-1} \mid g^n)\), which is crucial for accurately capturing the variability and intricacies of the transition between different diffusion stages. For our model, we set \(\sigma_\theta := \tilde{\beta}^n\), where \(\tilde{\beta}^n\) is a predetermined scaling factor that adjusts the noise level during each diffusion step, allowing for a controlled smoothing and detail enhancement in the generation process.

\begin{algorithm}[htbp]
\SetKwFor{For}{for}{do}{end\enspace for}
\SetKwIF{If}{ElseIf}{Else}{if}{then}{else if}{else}{end\enspace if}
\SetKw{Return}{Return:}
\caption{Sampling $g_{1:T}^0$ via annealed Langevin dynamics}
\KwIn{ noise $g_{1:T}^N \sim \mathcal{N}(0,I)$ and raw audio waveform \(a_{1:T}\)} 
\For {$n = N$ \emph{\KwTo} $1$}{
	\eIf{$n>1$}{$z \sim \mathcal{N}(0,I)$}{$z = 0$}
	$g_{1:T}^{n-1}=\frac{1}{\sqrt{\alpha^n}}\left(g_{1:T}^n - \frac{\beta^n}{\sqrt{1-\overline{\alpha}^n}}\epsilon_\theta\left(g_{1:T}^n,a_{1:T},n\right)\right)+\sqrt{\sigma_\theta}z$
}
\Return{$g^0_{1:T}$}
\label{alg:Inference}
\end{algorithm}

During inference, we send the entire sequence of raw audio to the condition extractor component. The output of this component is then fed to the diffusion model to generate the whole sequence of accompanying gestures (\(g^0\)).

\section{EXPERIMENTS}\label{sec:EXPERIMENTS}
To validate our approach, we utilized two co-speech gesture datasets: ZEGGS \cite{ghorbani2023ZeroEGGS} and BEAT \cite{liu2022BEAT}. Our experiments focused on producing full 3D body gestures, including finger motions and locomotion.

\subsection{Dataset and Data Processing}

\subsubsection{Datasets}
The ZEGGS dataset features a comprehensive collection of emotional expressions, enabling the exploration of gesture generation across a spectrum of emotions. Conversely, the BEAT dataset is distinguished by its focus on personalized movements, capturing the unique gesture styles of various individuals. This diversity in datasets allows for robust testing and enhancement of models aimed at generating nuanced and contextually appropriate co-speech gestures.

\subsubsection{Speech Audio Data Process}
Due to the WavLM large model being pre-trained on speech audio sampled at 16 kHz, we uniformly resampled all audio from the ZEGGS and BEAT datasets down from 44.1 kHz to match this frequency, ensuring compatibility and optimal performance.

\subsubsection{Gesture Data Process}
We concentrate exclusively on full-body gestures, employing the data processing techniques detailed by Alexanderson et al. \cite{alexanderson_simon_style-controllable_2020}. This includes capturing translational and rotational velocities to accurately delineate the root's trajectory and orientation. The datasets are uniformly downsampled to a frame rate of 20 fps. To ensure precise and continuous representation of joint angles, we utilize the exponential map technique \cite{grassiaf.sebastian1998Practical}. All data are segmented into 20-second clips for training and validation purposes. For user evaluation, the generated gesture sequences are divided into 10-second clips to enhance evaluation efficiency.

\subsection{Model Settings}
Our experiments utilize three AdaLN Mamba-2 blocks, with each Mamba-2 configured with a 256 SSM state expansion factor, a local convolution width of 4, and a block expansion factor of 2. This encoding process transforms each frame of the gesture sequence into hidden states \(h \in \mathbb{R}^{1280}\). We employ the pre-trained WavLM Large model\footnote{https://github.com/microsoft/unilm/tree/master/wavlm} for audio processing.

The diffusion model uses a quaternary variance schedule, starting from \(\beta_1 = 1 \times 10^{-4}\) to \(\beta_N = 8 \times 10^{-2}\) with a linear beat schedule, and a total of \(N = 1000\) diffusion steps. The training batch size is set to 32 per GPU.

Our model was tested on an Intel i9 processor with a 4090 GPU, in contrast to the A100 GPU used by Persona-Gestor. The training times were approximately 6 hours for ZEGGS and 22 hours for BEAT.

\subsection{Visualization Results}
Our system excels in generating personalized gestures that are contextually aligned with speech. By leveraging the Mamba-based fuzzy inference strategy, it autonomously derives fuzzy features directly from speech audio. The results demonstrate that our AdaLN Mamba-2 framework produces gesture actions comparable to those generated by Persona-Gestor, which uses convolution global fuzzy feature extractor and AdaLN transformers. 

\begin{figure}[htbp]
    \centering
    \subfloat[Happy]{\includegraphics[width=0.48\textwidth]{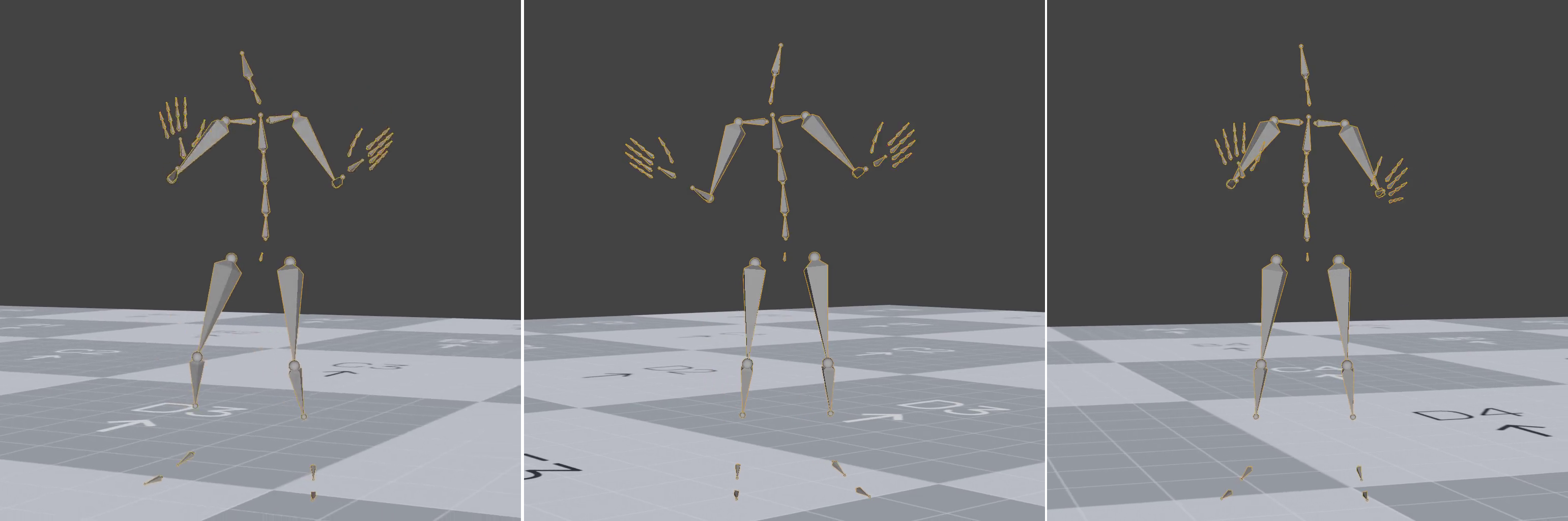}%
    \label{fig:happy} }
    \hfil
    \subfloat[Sad]{\includegraphics[width=0.48\textwidth]{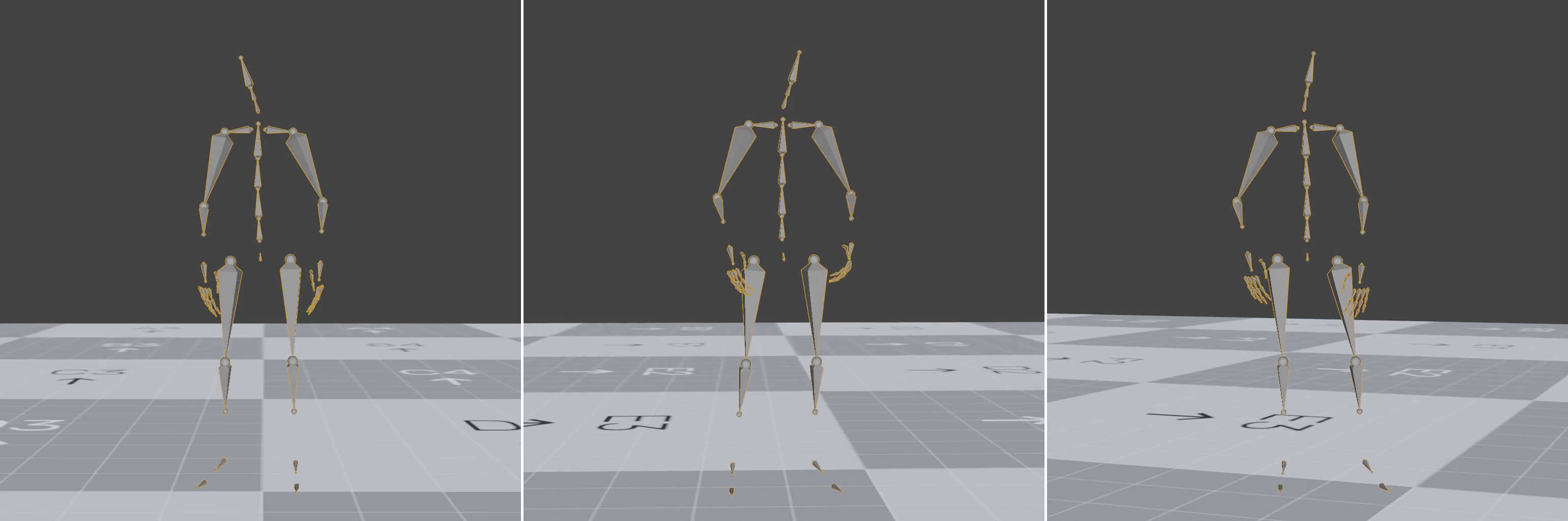}%
    \label{fig:sad} }
    \hfil
    \subfloat[Angry]{\includegraphics[width=0.48\textwidth]{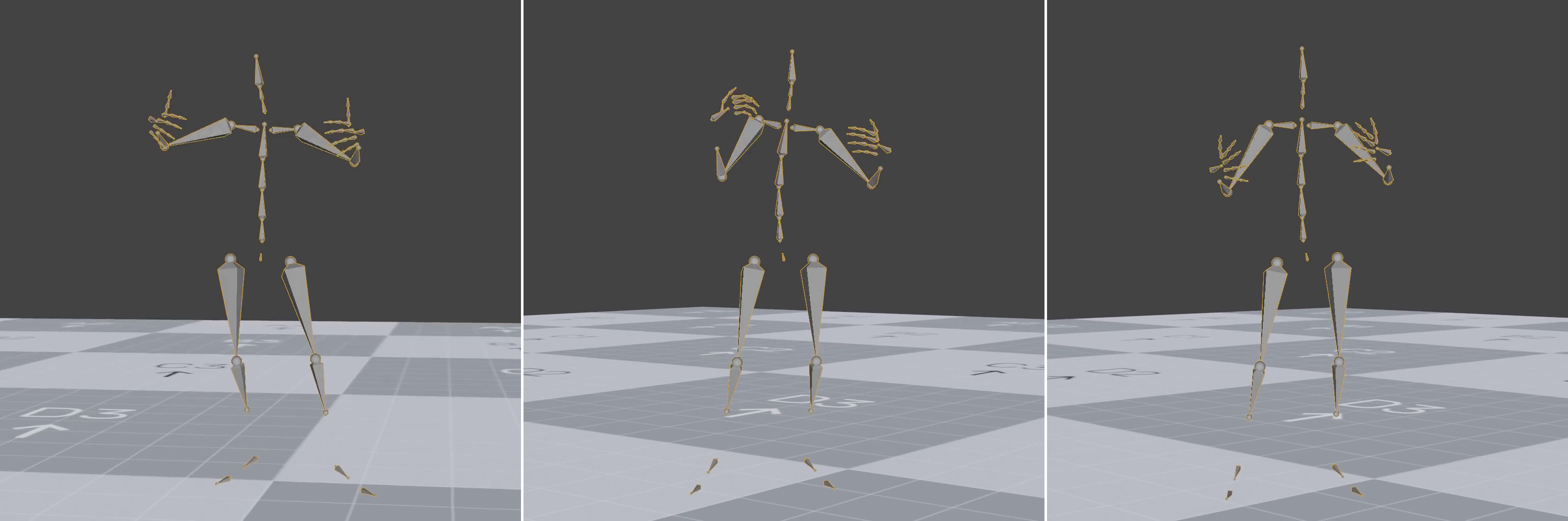}%
    \label{fig:angry} }
    \hfil
    \subfloat[Tired]{\includegraphics[width=0.48\textwidth]{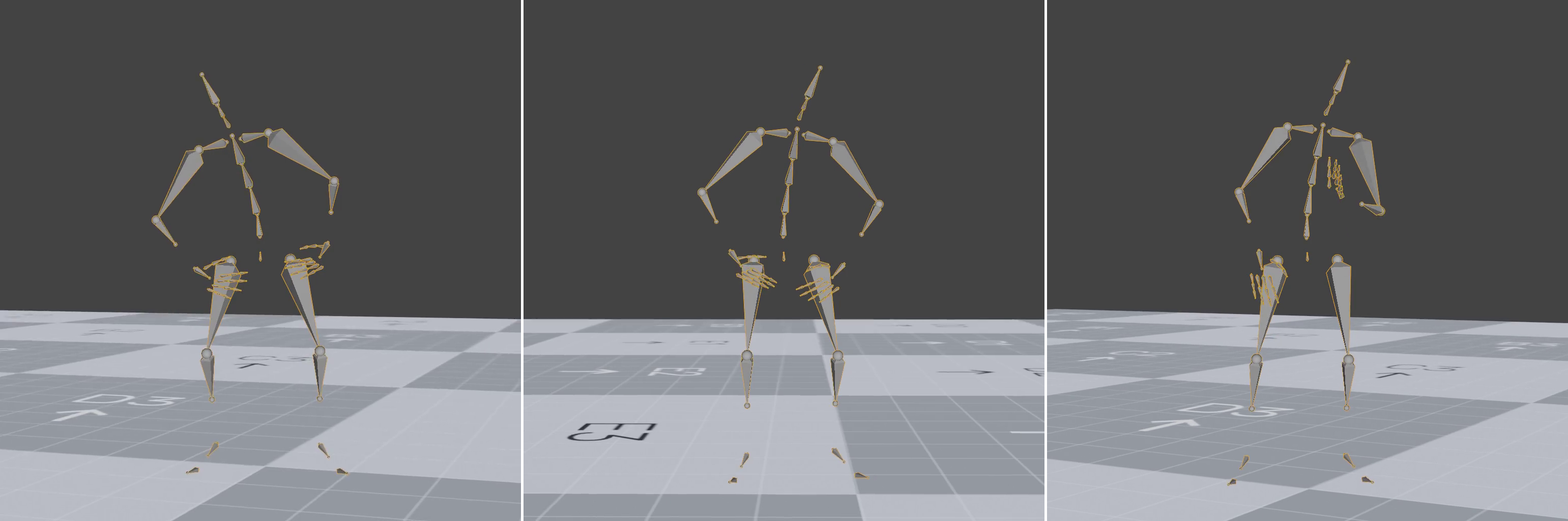}%
    \label{fig:tired} }
    \hfil
    \subfloat[old]{\includegraphics[width=0.48\textwidth]{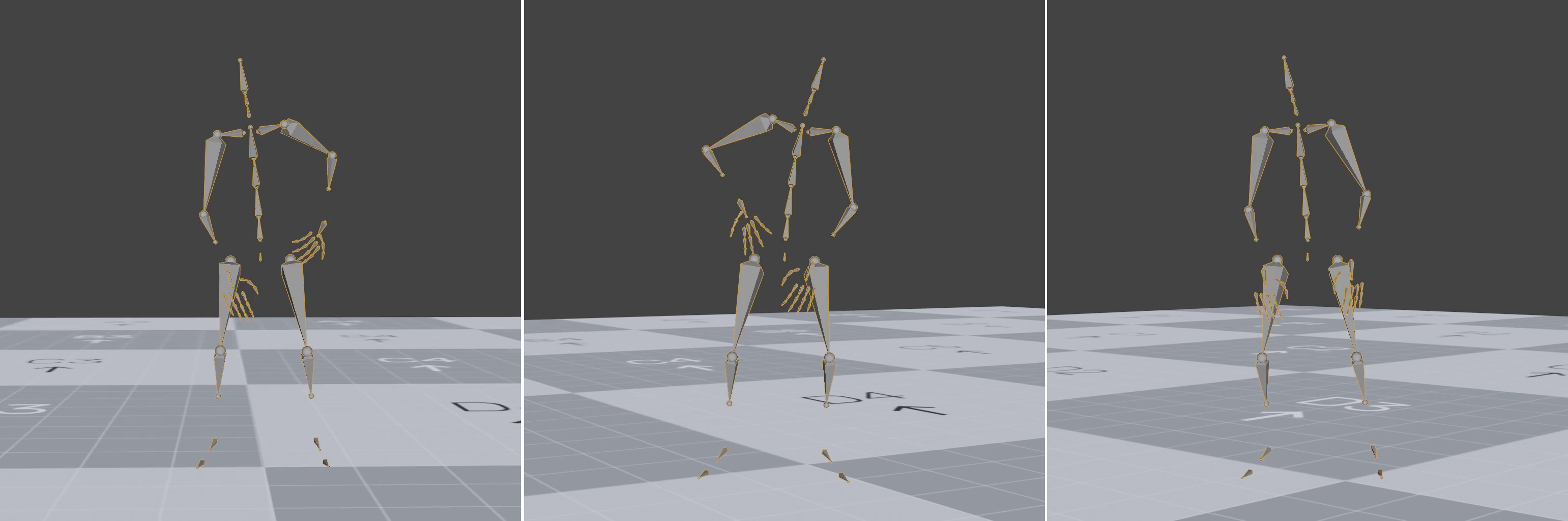}%
    \label{fig:old} }
    \hfil
    \subfloat[speech]{\includegraphics[width=0.48\textwidth]{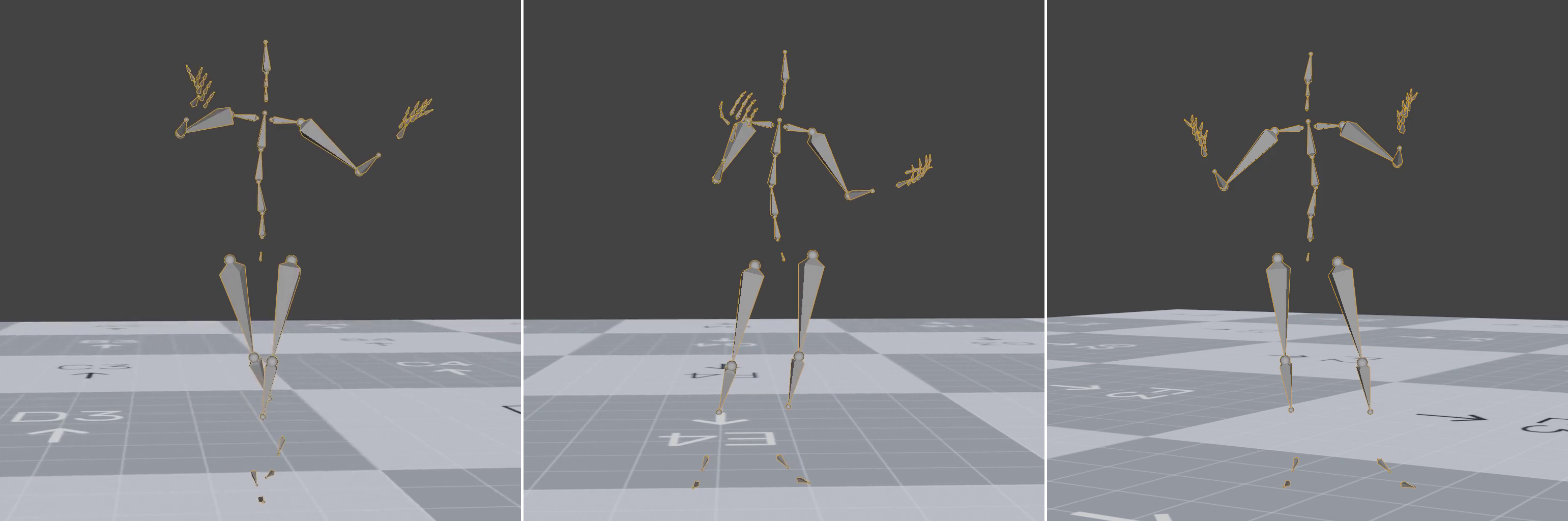}%
    \label{fig:speech} }
    \caption{Samples of gestures corresponding to various emotions are presented in the subfigure. The left side illustrates the ground truth gestures, the middle section showcases gestures generated by our architecture (DiM-Gesture), and the right side presents synthesized gestures using Persona-Gestor.}
    \label{fig:Emotion}
\end{figure}

Figure \ref{fig:Emotion} illustrates the visual outcomes of gestures aligned with the emotional valence conveyed by the audio. For example, the system generates gestures of joy in response to happy audio cues (see Figure \ref{fig:happy}) and gestures of sadness for sorrowful audio (as shown in Figure \ref{fig:sad}). Additionally, the system can infer age-related characteristics or other nuanced states from the speech audio (as depicted in \ref{fig:old} and \ref{fig:speech}). These visual comparisons demonstrate that DiM-Gesture can produce co-speech gestures of comparable quality to those generated by Persona-Gestor. 

\subsection{Subjective and Objective Evaluation}
In line with established practices in gesture generation research, we conducted a series of subjective and objective evaluations to assess the co-speech gestures generated by our proposed DiM-Gesture (DiM) model.

We adopted slightly varied baselines for different datasets. For the ZEGGS dataset, we employed LDA\cite{alexanderson2023listen}, and DiffuseStyleGesture (DSG)\cite{yang_diffusestylegesture_2023}. For the BEAT dataset, we utilized the same baseline models as in ZEGGS but replaced DSG with DSG+\cite{yang2023DiffuseStyleGestureaplus} and introduced GestureDiffuCLIP (GDC)\cite{ao2023GestureDiffuCLIP} as an additional baseline model. All baseline models employed are based on transformer and diffusion architectures.

In our experiments with the ZEGGS and BEAT datasets, we extended the original LDA, DSG, and DSG+ models to encompass all styles within these datasets. Efforts to adapt LDA to include finger motions encountered significant challenges, resulting in unsatisfactory gesture-generation outcomes. Consequently, we utilized gestures generated by the LDA model, excluding finger movements, for our analysis. 

\begin{table*}[ht]
\caption{The subject mean perceptual rating score. Bold fonts were utilized to emphasize the best results for each metric among the different methods, except for the GT.}
\label{tab:overview}
\centering
\resizebox{\linewidth}{!}{
\begin{tabular}{cccccccc}
\hline
                       & Methods                                                               & \multicolumn{3}{c}{Subject Evaluation Metric}                                                                                                      & \multicolumn{3}{c}{Objective Evaluation Metric}                                                                                                    \\ \hline
Dataset                & Model                                                                 & \begin{tabular}[c]{@{}c@{}}Human↑\\ likeness\end{tabular} & Appropriateness↑    & \begin{tabular}[c]{@{}c@{}}Style↑\\ appropriateness\end{tabular} & \begin{tabular}[c]{@{}c@{}}FGD↓\\ on feature space\end{tabular} & \begin{tabular}[c]{@{}c@{}}FGD↓\\ on raw data space\end{tabular} & BeatAlign↑    \\ \hline
\multirow{7}{*}{ZEGGS} & GT                                                                    & 0.93±1.22                                                 & 1.21±1.25           & /                                                                & /                                                               & /                                                                & /             \\
                       & LDA                                                                   & -0.76±1.28                                                & 0.013±1.35          & 0.38±1.87                                                        & 124.55                                                          & 50996.33                                                         & 0.66          \\
                       & DSG                                                                   & -0.42±1.37                                                & -0.76±1.08          & -0.79±1.18                                                       & 66.77                                                           & 33297.50                                                         & 0.63          \\
                       & PG                                                                    & \textbf{0.47±1.17}                                        & 0.476±1.29          & 0.76±1.52                                                        & 28.17                                                           & 26193.92                                                         & \textbf{0.67} \\
                       & \begin{tabular}[c]{@{}c@{}}(Proposed)\\ DiM-AdaLN-Mamba2\end{tabular} & 0.46±1.07                                                 & \textbf{0.478±1.32} & \textbf{0.77±1.35}                                               & \textbf{28.16}                                                  & \textbf{26013.12}                                                & \textbf{0.67} \\
                       & DiM-AdaLN-Mamba1                                                      & 0.21±1.23                                                 & 0.416±1.18          & 0.63±1.53                                                        & 35.21                                                           & 30453.23                                                         & 0.66          \\
                       & DiM-ConvSE                                                            & 0.45±1.27                                                 & 0.477±1.14          & 0.77±1.27                                                        & 28.53                                                           & 26878.34                                                         & \textbf{0.67} \\ \hline
\multirow{8}{*}{BEAT}  & GT                                                                    & 0.65±1.16                                                 & 0.96±1.04           & /                                                                & /                                                               & /                                                                & /             \\
                       & LDA                                                                   & -1.65±0.73                                                & -1.59±0.74          & -1.35±1.05                                                       & 264.06                                                          & 3471.26                                                          & 0.66          \\
                       & DSG                                                                   & -0.28±1.17                                                & -0.49±1.15          & -0.40±1.24                                                       & 23811.46                                                        & 2384465.64                                                       & 0.43          \\
                       & GDC                                                                   & 0.54±1.12                                                 & 0.47±1.25           & 0.30±1.27                                                        & 432.15                                                          & 93215.56                                                         & \textbf{0.69} \\
                       & PG                                                                    & 0.56±1.26                                                 & 0.64±1.07           & 0.66±1.33                                                        & \textbf{276.25}                                                 & \textbf{3584.95}                                                 & 0.68          \\
                       & \begin{tabular}[c]{@{}c@{}}(Proposed)\\ DiM-AdaLN-Mamba2\end{tabular} & \textbf{0.57±1.04}                                        & \textbf{0.65±1.15}  & \textbf{0.67±1.28}                                               & 276.32                                                          & 3607.75                                                          & 0.68          \\
                       & DiM-AdaLN-Mamba1                                                      & 0.37±1.15                                                 & 0.38±1.06           & 0.60±1.05                                                        & 283.24                                                          & 3924.58                                                          & 0.66          \\
                       & DiM-ConvSE                                                            & 0.56±1.22                                                 & \textbf{0.65±1.32}  & \textbf{0.67±1.38}                                               & 276.58                                                          & 3683.65                                                          & 0.68          \\ \hline
\end{tabular}
}
\end{table*}

\subsubsection{Subjective Evaluation}

For comprehensive subjective evaluations, we employ three metrics: human-likeness, appropriateness, and style-appropriateness. Human-likeness assesses the naturalness and resemblance of gestures to authentic human movements, independent of speech. Appropriateness evaluates the temporal alignment of gestures with speech rhythm, intonation, and semantics, ensuring a natural flow. Style-appropriateness measures the similarity between generated gestures and their original human counterparts.

We conducted a user study employing pairwise comparisons, as recommended by \cite{wolfert_rate_2021}. In each trial, participants were presented with two 10-second video clips side by side, generated by different models, including the Ground Truth (GT), for direct comparison. Participants were instructed to select the clip they preferred based on specified evaluation criteria. Preferences were quantified on a scale from 0 to 2, with the non-selected clip in each pair receiving an inverse score (e.g., a -2 score for the non-chosen clip if the chosen one received a score of 2). A score of zero indicated a neutral preference.

Given the extensive range of styles in the ZEGGS (19) and BEAT (30) datasets, evaluating each style individually was deemed impractical. Therefore, we employed a random selection method, assigning each participant a subset of five styles from ZEGGS and six characters from BEAT for evaluation. Importantly, none of the selected audio clips were included in the training or validation sets, ensuring the integrity of the assessments.

Thirty volunteer participants—14 males and 16 females aged between 18 and 35—were recruited for this study. All participants demonstrated a high level of English proficiency, essential for accurately interpreting and responding to the tasks involved.

One-way ANOVA and posthoc Tukey, multiple comparison tests, were conducted to determine if there were statistically significant differences among the models' scores across the three evaluation aspects. The results are presented in Table \ref{tab:overview}, offering detailed insights into the performance variances observed among different models regarding human-likeness, appropriateness, and style-appropriateness. 

The results from the ZEGGS and BEAT datasets show that the Ground Truth (GT) achieves the highest scores ($0.93\pm1.22$ for ZEGGS and $0.65\pm1.16$ for BEAT), exhibiting statistically significant differences ($p<0.001$) in human-likeness evaluations compared to model-generated gestures. The GT features a diverse yet limited selection of gestures, each characterized by distinct traits that enhance the realism of movements. However, these specific gestures fall into the long-tail distribution of the datasets, presenting substantial challenges to the learning capabilities of the models. Furthermore, the uniqueness of these gestures significantly influences the appropriateness and style-appropriateness scores. 

The evaluations of the ZEGGS and BEAT datasets showed no statistically significant differences ($p>0.05$) between our DiM model and Persona-Gestor (PG) across all three metrics. However, PG scored slightly higher in Human-likeness on the ZEGGS dataset, whereas DiM attained marginally higher scores in Appropriateness and Style-appropriateness. In contrast, on the BEAT dataset, DiM scored slightly higher across all subjective evaluation metrics. These outcomes demonstrate that DiM is capable of generating co-speech gestures of comparable quality to those produced by PG, affirming the robustness and effectiveness of our approach in synthesizing natural and well-aligned gestures across diverse datasets. 

\subsubsection{Objective Evaluation}\label{sec:objective eval}
We employ three objective evaluation metrics to assess the quality and synchronization of generated gestures: Fréchet Gesture Distance (FGD) in both feature and raw data spaces\cite{yoonyoungwoo2020Speecha}, and BeatAlign\cite{li2021ai}. Inspired by the Fréchet Inception Distance (FID)\cite{heusel2017gans}, FGD evaluates the quality of generated gestures and has shown moderate correlation with human-likeness ratings, surpassing other objective metrics\cite{kucherenko2023Evaluating}. BeatAlign, on the other hand, assesses gesture-audio synchrony by calculating the Chamfer Distance between audio beats and gesture beats, thus providing insights into the temporal alignment of gestures with speech rhythms.

Table \ref{tab:overview} presents our results, underscoring the state-of-the-art performance of our method in objective evaluations using the Fréchet Gesture Distance (FGD) metrics. Our model achieves superior performance (28.16 for ZEGGS) compared to other architectures, generating gestures that closely align with the Ground Truth (GT). It also matches the BeatAlign scores (0.67 for ZEGGS) of other models, except for GestureDiffuClip (GDC) which scores slightly higher (0.69 for BEAT), highlighting its efficacy in producing co-speech gestures that synchronize accurately with speech rhythms. Despite GDC's high BeatAlign score, corroborating user feedback indicates its overemphasis on prosodic cues results in frequent high-frequency gestures that, while technically accurate, reduce the naturalness of the gestures. On the BEAT dataset, our model's FGD scores (276.32) are marginally lower than Persona-Gestor (276.25), reflecting competitive performance in gesture quality.

\subsection{Ablation Studies}

Ablation studies were conducted to assess the influence of crucial components within our model, particularly focusing on the Mamba-based fuzzy feature extractor and the AdaLN Mamba-2 architecture. These studies aimed to elucidate how each component contributes to the overall performance and effectiveness of the gesture synthesis process. 
\subsubsection{Ablation of Mamba-based Style Fuzzy Feature Extractor}
For the Mamba-based style fuzzy feature extractor, we conducted an investigation into the implications of replacing the Mamba architecture with a convolutional style extractor (similar to the approach in PG), which we termed DiM-ConvSE. This experiment aimed to assess the differential impacts of the Mamba and convolutional methodologies on the effectiveness and fidelity of the synthesized gestures. 

The results, as detailed in Table \ref{tab:overview}, reveal no statistically significant differences ($p>0.05$) in the style-appropriateness metrics between DiM and DiM-ConvSE for both the ZEGGS and BEAT dataset experiments. These findings underscore the equivalence of the Mamba-based style fuzzy feature extractor in performance with the convolution-based extractor, highlighting its efficacy in gesture synthesis. Additionally, as indicated in Table \ref{tab:ablation}, the Mamba architecture demonstrates less parameter count and lower memory consumption than DiM-ConvSE.

\begin{table}[H]
\caption{In the ablation study focusing on the Mamba-based style fuzzy feature extractor and the AdaLN Mamba2, we meticulously assessed the parameter count and inference time to quantify the efficiency and performance impact of these components.}
\resizebox{\linewidth}{!}{
\label{tab:ablation}
\begin{tabular}{ccc}
\hline
                                                                      & Param. Count & \begin{tabular}[c]{@{}c@{}}Inference Time\\  (length of 80s Gesture Sequence)\end{tabular} \\ \hline
PG                                                                    & 1.2B         & 82s                                                                                       \\
\begin{tabular}[c]{@{}c@{}}(Proposed)\\ DiM-AdaLN-Mamba2\end{tabular} & 421M         & 30s                                                                                        \\
DiM-AdaLN-Mamba1                                                      & 426M         & 15s                                                                                        \\
DiM-ConvSE                                                            & 711M         & 9s                                                                                        \\ \hline
\end{tabular}
}
\end{table}

\subsubsection{Ablation of AdaLN Mamba-2}
In our ablation study, we explored the impact of different versions of the Mamba architecture by replacing the AdaLN Mamba-2 with Mamba-1 \cite{gu_mamba_2023}. This experiment was designed to assess the efficacy of the AdaLN Mamba-2 in handling the intricate dynamics of gesture generation compared to its predecessor, Mamba-1. The main focus was to determine whether the advancements in Mamba-2, particularly the integration of adaptive layer normalization (AdaLN), provide substantial improvements in gesture synthesis quality, efficiency, and model responsiveness. The results of this comparison are detailed in Table \ref{tab:overview} and Table \ref{tab:ablation}, illustrating differences in gesture quality, computational efficiency, and performance metrics between the two Mamba versions.

In our ablation studies on the AdaLN Mamba-2, substituting the Mamba-2 module with Mamba-1 led to a significant decline in performance across all metrics. This deterioration can be attributed to the earlier architecture's inability to precisely synchronize speech rhythm and capture stylistic nuances like its successor. These findings highlight the integrated enhancements in Mamba-2, particularly its advanced capability to effectively handle the intricate dynamics of speech-driven gesture generation. However, it is worth noting that DiM-AdaLN-Mamba-1 exhibits a faster inference time.

\section{DISSCUSTION and CONCLUSION}\label{sec:CONCLUSION}
In this study, we present \textit{DiM-Gesture}, an innovative network architecture designed to generate personality-specific gestures directly from raw speech audio using a Mamba-based architecture. \textit{DiM-Gesture} incorporates a Mamba-based style fuzzy feature extractor and an AdaLN Mamba-2 diffusion architecture, facilitating the seamless synthesis of nuanced, personality-driven gestures. Our approach achieves a quality of action generation comparable to that of the AdaLN Transformer architecture (Persona-Gestor) while requiring significantly less memory and substantially enhancing generation speed than Persona-Gestor. This demonstrates \textit{DiM-Gesture}'s capability to provide efficient and high-quality gesture synthesis.

The Mamba-based fuzzy feature extractor employs a fuzzy feature inference strategy within its dual-component module to autonomously infer both fuzzy stylistic features and specific audio details. These elements are merged into a unified latent representation, enabling the generation of speaker-aware personalized 3D full-body gestures. This approach integrates a pivotal innovation in synthesizing personality-driven gestures by leveraging automatically inferred fuzzy features, thereby eliminating the need for explicit style labels or additional features. Such advancements facilitate an end-to-end gesture generation process that authentically reflects the speaker's unique characteristics directly from raw speech audio. The integration of fuzzy feature inference streamlines the creation process, enhancing both generalization capabilities and user accessibility.

The AdaLN Mamba-based mechanism, a conditional architecture, uniformly applies a specific function across all sequence tokens, significantly enhancing the model's ability to capture and represent both conditional dependencies and output characteristics efficiently. Like AdaLN transformers, AdaLN Mamba-2 enhances the understanding and processing of the complex interactions between continuous fuzzy features as conditional inputs and resultant gesture synthesis. This leads to improved model performance and output fidelity. Ultimately, \textit{DiM-Gesture} employs a diffusion mechanism to produce a diverse spectrum of gesture outputs, showcasing its capability to handle varied and nuanced gesture synthesis effectively.

Our study highlights key areas for enhancement: There remains a noticeable disparity between the current batch generation form and the desired real-time generation capability. Bridging this gap is crucial for applications requiring immediate gesture synthesis, such as live interactions or performances.

\section{ACKNOWLEDGMENTS}
This work was partially supported by the "Pioneer" and "Leading Goose" R\&D Program of Zhejiang (No.2023 C01212), the National Key Research and Development Program of China (No.2022YFF 0902305), the Public Welfare Technology Application Research Project of Zhejiang (No.LGF21F020002, No.LGF22F020008), the Key Program and development projects of Zhejiang Province of China (No.2021C03137), and the Key Lab of Film and TV Media Technology of Zhejiang Province (No.2020E10015).

\bibliography{mybibliography}

\bibliographystyle{IEEEtran}

\end{document}